\documentclass{acm_proc_article-sp}
\usepackage{psfrag}

\begin{document}

\title{Combined Intra- and Inter-domain Traffic Engineering using Hot-Potato Aware Link Weights Optimization}

\numberofauthors{1}
\author{
\alignauthor Simon Balon\titlenote{S. Balon is Research Fellow of the Belgian National Fund for the Scientific Research (FNRS) and also partially funded by the EU under the ANA FET project (FP6-IST-27489).} and Guy Leduc\\
       \affaddr{Research Unit in Networking}\\
       \affaddr{Universit\'e de Li\`ege (ULg) - Belgium}\\
       \email{\{balon, leduc\}@run.montefiore.ulg.ac.be}
}

\maketitle
\begin{abstract}

A well-known approach to intradomain traffic engineering consists in
finding the set of link weights that minimizes a network-wide
objective function for a given intradomain traffic matrix. This
approach is inadequate because it ignores a potential
impact on interdomain routing. Indeed, the resulting set of link
weights may trigger BGP to change the BGP next hop for some
destination prefixes, to enforce hot-potato routing policies. In turn,
this results in changes in the intradomain traffic matrix that have
not been anticipated by the link weights optimizer, possibly leading
to degraded network performance.

We propose a BGP-aware link weights optimization method that takes these effects
into account, and even turns them into an advantage. This method uses
the interdomain traffic matrix and other available BGP data, to extend
the intradomain topology with external virtual nodes and links, on
which all the well-tuned heuristics of a classical link weights
optimizer can be applied. A key innovative asset of our method is its
ability to also optimize the traffic on the interdomain peering
links. We show, using an operational network as a case study, that our
approach does so efficiently at almost no extra computational cost.

\end{abstract}

\section{Introduction \& Motivation}
\label{sec:intro}

Intradomain traffic engineering consists in routing traffic in an
optimal way from ingress nodes to egress nodes in a given domain. If
shortest path IP routing is used, the only way to optimize the traffic
is by finding an appropriate set of link weights that minimizes a
given domain-wide objective function. For example, if this objective
is to minimize the total (equivalently the average) link \textit{load}, a
solution consists in assigning unitary weights to all links in the
network. On the other hand, to minimize the average link
\textit{utilization}, the solution consists in choosing link
weights that are inversely proportional to the link
capacities\footnote{In \cite{balon-icon06} we demonstrate that these
  link weights settings minimizes these objective functions.}. Note
that these two examples are representative of traffic-independent link weights
settings, i.e. they minimize their respective objectives for \textit{every} possible
traffic matrix. 

For other objective functions (e.g. minimizing the maximum link load
or utilization), the optimal choice of link weights usually depends on
the traffic matrix. Therefore in its simplest form the resolution of
this optimization problem needs to take as inputs (1) the network
topology with unknown link weights, (2) the chosen network-wide objective
function, and (3) an intradomain traffic matrix, which specifies the
amount of traffic between every pair of ingress/egress nodes
\cite{fortz1}. This optimization problem is NP-hard and good
local-search heuristics are thus needed to find a set of link weights that
reasonably minimizes the objective function in a reasonable time. 

However this approach is unaware of the interdependence between
intradomain and interdomain routings. Actually the real traffic demand
is an interdomain traffic matrix (from prefix to prefix), while the
intradomain traffic matrix (from ingress to egress nodes) is only the
result of applying BGP routing decisions on the interdomain traffic
matrix (TM). Even if we consider that the interdomain TM and the
interdomain (BGP) routes are invariant, the intradomain TM may still
vary if some link weights are changed inside the domain. This is due to
the so-called hot-potato (or early exit) decision rule implemented by BGP. 

\begin{figure}[htbp]
  \centering
  \includegraphics[width=6cm]{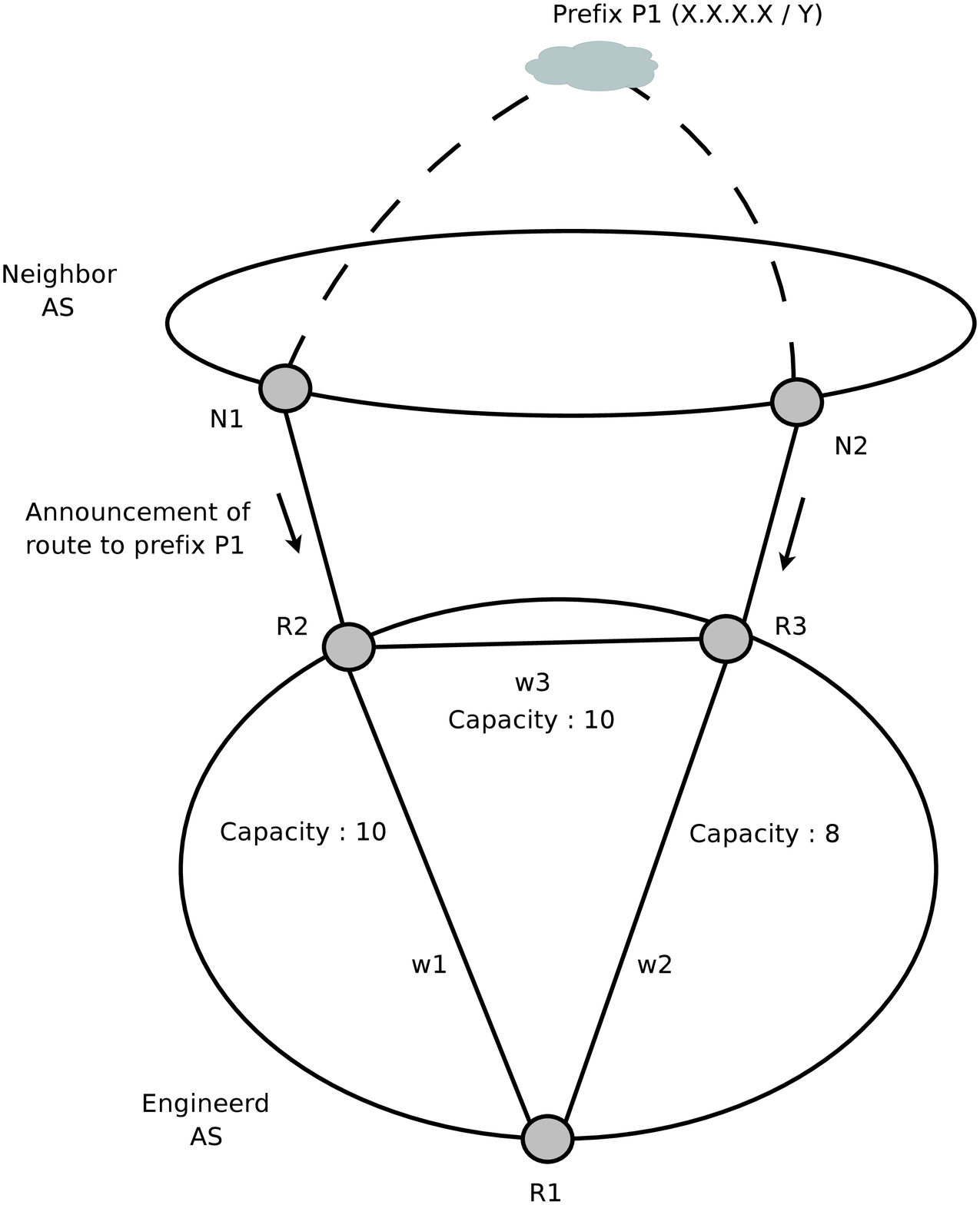}
  \caption{Toy Example}
  \label{fig:toyexample}
\end{figure}

The toy example depicted in figure \ref{fig:toyexample} suffices to
illustrate the problem\footnote{This example network is similar to the
one used in \cite{selin}}. This figure shows a domain with three nodes:
an ingress node $R_1$ possibly sending traffic to egress nodes $R_2$ and
$R_3$, and three intradomain links of weights $w_1$, $w_2$ and $w_3$. Suppose
this domain (also called an AS) has two peering links (respectively
$R_2$-$N_1$ and $R_3$-$N_2$) with a neighboring AS providing connectivity to
the IP prefix $P_1$. Further suppose that no BGP rule of higher
precedence than the hot-potato rule has been able to make a selection
between $R_2$ and $R_3$. If the link weights are inversely proportional to
the link capacities shown on the figure, then ingress node $R_1$ will
choose to reach this prefix through egress node $R_2$ according to the
hot-potato rule (because $w_1$ = 1/10 $<$ $w_2$ = 1/8). If $R_1$ has 5 units
of traffic to send to $P_1$, then the intradomain TM is just 5 units
from $R_1$ to $R_2$ and no traffic elsewhere. 

Now suppose that we run a link weights optimizer (denoted LWO in the sequel) that tries to
minimize the maximum link utilization, while allowing equal cost
multipath (ECMP)\cite{OSPF}. A possible optimal link weights setting
is {$w_1$ = 2, $w_2$ = $w_3$ = 1}, leading to two IGP equal cost paths from
$R_1$ to $R_2$ and to a maximum link utilization of 2.5/8 on link
$R_1$-$R_3$. However, if the weights are set as proposed, the hot-potato
rule will now select $R_3$ as egress node to reach $P_1$ (because $w_2$ $<$
$w_1$), and the resulting intradomain TM is actually 5 units of traffic
from $R_1$ to $R_3$, with a maximum utilization of 5/8 on link
$R_1$-$R_3$. Clearly, the outcome is much worse than expected, and even
worse than keeping the initial weights setting! 

This toy example illustrates that we cannot rely on the intradomain TM
to solve the optimization problem, because it is not invariant under
link weights changes, possibly leading to degraded network
performance. 

Even though this toy example is not representative of real networks
with real traffic, we will show in section \ref{sec:casestudy}, by using an operational
network as a case study, that this phenomenon can really happen with
bad consequences, because a substantial
amount of prefixes/traffic may be subject to hot-potato (re)routing. 
For the case study in section \ref{sec:casestudy} we show that 97.2\% of the prefixes
have multiple possible egress points, which amounts to 35.6\% of the traffic
on average. Without taking hot-potato effects into account, we will show that the link
weights proposed by a classical LWO 
may result in link utilizations close to and even above 100\%,
while the tool expected maximal 
link utilizations of only about 35\%.

We propose a link weights optimization method that takes
these hot-potato effects into account, and even turns them into an
advantage. To this end we use as inputs the (hot-potato invariant)
interdomain TM and some BGP data, both collected inside the domain, to
infer the set of IP prefixes that can be reached by at least two
egress nodes and for which no BGP rule of higher precedence than the
hot-potato rule has been able to make a selection, i.e. for which the
hot-potato rule can potentially be the tie-breaker. We call this
subset the hot-potato prefixes, and from now on in this introduction
we will only consider these prefixes. 

\begin{figure}[htbp]
  \centering
  \includegraphics[width=3cm]{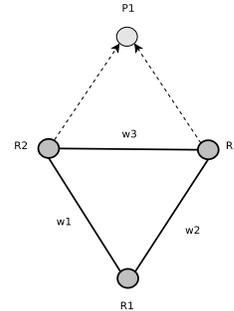}
  \caption{Toy Example - Simplified Version}
  \label{fig:toyexample-simplified}
\end{figure}

Our method is based on an extension of the intradomain topology with
external virtual nodes and links. A first naive and unscalable way to
solve the problem would consist in adding a virtual node per
hot-potato prefix and attach this node to all possible BGP next-hops
for this prefix. This is depicted on figure \ref{fig:toyexample-simplified} for the toy example
of figure \ref{fig:toyexample}. If we now run LWO on this virtual
topology, while still allowing equal cost multipaths, including
multiple BGP next-hops, an optimal weights setting is {$w_1$ = $w_2$ = $w_3$
  = 1}, which will split the 5 units of traffic evenly on the two
paths $R_1$-$R_2$-$N_1$ and $R_1$-$R_3$-$N_2$. 

However, the number of hot-potato prefixes can be very large and we
would like to keep the number of virtual nodes roughly similar to the
number of ordinary nodes. To this end we propose to aggregate all
virtual nodes attached to exactly the same sets of BGP next-hops. They
are indeed indistinguishable with respect to intradomain routing. On
the operational network that we have considered, the number of such
nodes boils down from 160,000 to only 26. We further show by
considering the amount of traffic sent to these aggregates, that we
can reduce this set to only 5 virtual nodes without losing more than
0.06\% of the total "hot-potato" traffic. 

An asset of our method lies in reusing the well-tuned LWO
heuristics on this extended topology. Moreover, we have also extended
this intradomain traffic engineering problem to the
peering links, by taking these links into account in the objective
function. In our simulations this method allowed us to reduce the maximal
interdomain link utilization from 70.1\% to 36.5\%.

The paper is structured as follows. In section \ref{sec:relatedwork} we review related
works. In section \ref{sec:survey} we present the necessary knowledge on intra- 
and interdomain routings. In section \ref{sec:method} we formulate the problem and propose our
BGP-aware LWO. In section \ref{sec:casestudy} we show an application of
the method using an operational network. In section \ref{sec:extensions}, we discuss
future work concerning potential oscillations. Finally, section \ref{sec:conclusion}
concludes the paper.

\section{Related Work}
\label{sec:relatedwork}

A first LWO algorithm for a given intradomain traffic matrix has been 
proposed by Fortz et al. in \cite{fortz1}.
It is based on a tabu-search metaheuristic and finds a
nearly-optimal set of link weights that minimizes a particular objective function, 
namely the sum over all links of a convex function of the link loads and/or utilizations. 
This problem has later been generalized to take several
traffic matrices \cite{fortz2} and some link failures \cite{fortz3} into
account. 
A heuristic that takes into account possible link failure
scenarios when choosing weights is also proposed in \cite{nucci} by Nucci et al. 
In our LWO we reuse the heuristic detailed in \cite{fortz1}, but 
we have adapted this algorithm to consider the effect of hot-potato routing. 
All the later improvements to this algorithm
(i.e. multiple traffic matrices, link failures) could be
integrated in our new LWO in a similar way.

The fact that the intradomain TM is not the correct input for many Traffic Engineering
problems had already been pointed out in  \cite{feldmann,rexford2} by Feldmann et al., 
who suggested to consider the set of possible
egress links in the traffic matrix. 
In \cite{rexford1} several extensions to the classical LWO problem are briefly described by Rexford, 
including a sketch of a method that resembles ours. 
Our work is in line with this recommendation, 
as we connect several equivalent egress nodes to a single virtual node representing the destination,
but our paper proposes a complete method to solve the link weights optimization problem, 
applicable to intradomain and peering links, 
and we demonstrate its efficiency on an operational network.
In \cite{feldmann} some methods to compute traffic matrices from
netflow traces are also presented, which are reused in this paper.
In \cite{agarwal05} Agarwal et al. study how hot-potato routing
  influences the selection of IGP link metrics and how traffic to
  neighboring ASes shifts due to changes in the
  local AS's link metrics. In their measurement study they find that
  metrics resulting from ignoring hot-potato interaction can be
  sub-optimal by as much as 20\% of link utilization. We show in this
  paper that the sub-optimality can be much larger. They also find
  that as much as 25\% of traffic to a neighboring AS can shift the
  exit point due to a local AS IGP link metric optimization. They have
  developed a patch to their link weight optimizer which recomputes
  the intradomain traffic matrix from the interdomain one at each step
  of the optimization. Their optimizer does not consider directly the
  interdomain traffic matrix so nothing prevents it to indefinitely
  iterate, as it would probably do in the toy example of figure \ref{fig:toyexample} if the
  heuristic tries to tune the link weights to enable intradomain
  equal cost multipaths. Also, their link weights optimizer does not
  engineer interdomain links and they have tested their algorithm on
  only 80\% of the total traffic of their private ISP while we have tested it on 99\% of
  traffic of an operational network. 
  Finally their source code is not available, while our algorithm 
  will be available in open-source in the TOTEM toolbox (\cite{totem}).

Cerav-Erbas et al. have already shown in \cite{selin} that the link
weights found by a LWO may change the intradomain TM considered as
input. In that paper they also show that applying LWO recursively on the resulting intradomain TM may not converge.
They propose a method that keeps track of the series of resulting TMs 
and at each iteration they optimize the weights for \textit{all} the
previous resulting intradomain TMs simultaneously.
However, they do not consider the general problem with
multiple exit points for each destination prefix, let alone taking
advantage of it.

In \cite{cope} a class of traffic engineering
algorithms is proposed by Wang et al. to optimize for the expected scenarios while providing a
worst-case guarantee for unexpected scenarios. They propose to
take the interdomain routing into account by splitting the problem into
two subproblems. The first one consists in optimizing the mapping of every (hot-potato) 
destination prefix to a single egress point. 
This can then be implemented in BGP by assigning a higher local preference 
to the route received by the chosen egress node.
The second subproblem is then the classical link weights optimization 
for the resulting (and now invariant) intradomain TM. In our approach we solve 
both subproblems in one step with the usual LWO and
we do not necessarily need to assign local preference values to
pin down every destination prefix to a unique BGP next-hop. 
By keeping all the potential next-hops we have more flexibility to engineer the network.

Several studies have shown that the proportion of prefixes whose next hop is
selected by the hot-potato criterion can be very large in ISP networks.
Based on measurements of one ISP network (AT\&T's tier-1 backbone
network) Teixeira et al. show in \cite{teixeira1} that
hot-potato routing changes are responsible for a big part of BGP
routing changes. While this is not the main goal of that paper they
have measured that more than 60\% of the prefixes can be affected by
the hot-potato routing changes and that these {\it hot-potato}
prefixes account for 5-35\% of the traffic in the network.
It is also explained in \cite{teixeira3} that "Since large ISPs
typically peer with each other in multiple locations, the hot-potato
tie-breaking step almost always drives the final routing decision for
destinations learned from peers, although this is much less common for
destinations advertised by customers.". The authors show that although
most routing changes do not cause important traffic shifts, routing is
a major contributor to large traffic variations. This demonstrates that
it is very important to take BGP routing considerations into account
when running traffic engineering algorithms.

In \cite{roughan2003} Roughan et al. analyse the effects of
  imprecision in the traffic matrix due to estimation techniques on traffic
  engineering algorithms. While the effects of these imprecisions
  seem to be quite limited, we show in this paper that the
  effects due to hot-potato routing can be very large. This is an
  important result as this highlights that not taking hot-potato
  effects into account cannot be simply seen as resulting in little
  (harmless) imprecision in the traffic matrix. Hot-potato errors
  in the TM can really be a big problem for intradomain TM-based TE
  algorithms optimizing the link weights.

An important point in the whole traffic engineering process is the
selection of a (set of) traffic matrix(ces) to use as input of the
traffic engineering algorithm. This problem is addressed in \cite{zhang} by Zhang and Ge, who try and
find such a subset of critical traffic matrices from the whole set of
measured traffic matrices. This work is complementary to ours.

To the best of our knowledge this paper proposes the first algorithm to find the
best possible set of link weights to engineer intra- and
  inter-domain links while taking hot-potato effects into account. 

\section{Routing principles}
\label{sec:survey}

\bigskip

\begin{figure}[htbp]
  \psfrag{P1}{{\scriptsize $P_1$}}
  \psfrag{P2}{{\scriptsize $P_2$}}
  \psfrag{P3}{{\scriptsize $P_3$}}
  \psfrag{P4}{{\scriptsize $P_4$}}
  \psfrag{N1}{{\scriptsize $N_1$}}
  \psfrag{N2}{{\scriptsize $N_2$}}
  \psfrag{N3}{{\scriptsize $N_3$}}
  \psfrag{N4}{{\scriptsize $N_4$}}
  \psfrag{R1}{{\scriptsize $R_1$}}
  \psfrag{R2}{{\scriptsize $R_2$}}
  \psfrag{R3}{{\scriptsize $R_3$}}
  \psfrag{R0}{{\scriptsize $R_0$}}
  \psfrag{AS 1}{{\scriptsize $AS1$}}
  \psfrag{AS 2}{{\scriptsize $AS2$}}
  \psfrag{1/1}{{\scriptsize $1/1$}}
  \psfrag{1/2}{{\scriptsize $1/2$}}
  \psfrag{1/3}{{\scriptsize $1/3$}}
  \psfrag{1/4}{{\scriptsize $1/4$ }}
  \psfrag{Engineered AS}{{\scriptsize Engineered AS}}
  \centering
  \includegraphics[width=4.5cm]{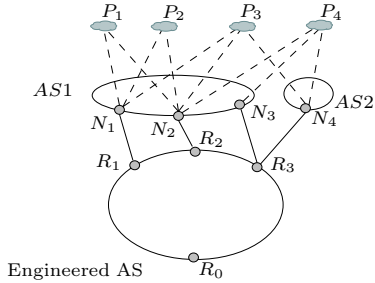}
  \caption{More Complex Topology}
  \label{fig:example2nonodes}
\end{figure}

Each packet sent on the Internet follows a path which is defined by
routing protocols. The exterior gateway
protocol (EGP) defines the path at the network-level. This path is called
the AS path\footnote{AS stands for Autonomous System. In the paper we use domain
  and AS interchangeably.}. The EGP used
in the Internet is BGP (Border Gateway Protocol). In each AS the path
from each ingress router to each egress router is defined by the
interior gateway protocol (IGP). The IGPs most commonly used in
transit networks are OSPF and ISIS. 

\begin{figure}[htbp]
  \psfrag{P1}{{\scriptsize $P_1$}}
  \psfrag{P2}{{\scriptsize $P_2$}}
  \psfrag{P3}{{\scriptsize $P_3$}}
  \psfrag{P4}{{\scriptsize $P_4$}}
  \psfrag{N1}{{\scriptsize $$}}
  \psfrag{N2}{{\scriptsize $$}}
  \psfrag{N3}{{\scriptsize $$}}
  \psfrag{N4}{{\scriptsize $$}}
  \psfrag{R1}{{\scriptsize $R_1$}}
  \psfrag{R2}{{\scriptsize $$}}
  \psfrag{R3}{{\scriptsize $$}}
  \psfrag{R0}{{\scriptsize $R_0$}}
  \psfrag{AS 1}{{\scriptsize $AS1$}}
  \psfrag{AS 2}{{\scriptsize $AS2$}}
  \psfrag{1/1}{{\scriptsize $1/1$}}
  \psfrag{1/2}{{\scriptsize $1/2$}}
  \psfrag{1/3}{{\scriptsize $1/3$}}
  \psfrag{1/4}{{\scriptsize $1/4$ }}
  \psfrag{Engineered AS}{{\scriptsize Engineered AS}}
  \centering
  \includegraphics[width=4.5cm]{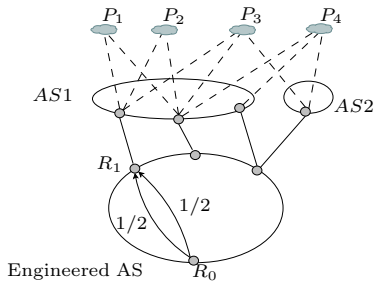}
  \caption{Intradomain Equal Cost Multipath (ECMP)}
  \label{fig:example2ecmp}
\end{figure}

In an AS the path between ingress and egress routers are
computed by a Shortest-Path algorithm based on the
link weights. If ECMP (Equal Cost Multi-Path) is enabled, several equal shortest-paths
can be used simultaneously to evenly
split the traffic among them, by using a hash table that
maps a hash of multiple fields in the packet header to one of these paths, 
so that all packets of a flow
will follow the same path with limited packet reordering (see
\cite{cao} for a performance analysis of hashing based schemes for
Internet load balancing). Figure \ref{fig:example2ecmp} shows an
example of ECMP inside an AS. This figure assumes that there are 
two equal cost paths from $R_0$ to $R_1$.

BGP allows routers to exchange reachability
information between neighboring ASes (\cite{bgp}). Each AS is
connected to several neighboring ASes by interdomain
links. Depending on the connectivity of the network and on the
destination of the packet, one or several neighboring ASes can be
chosen to forward the packet to the destination. The choice of the BGP next-hop 
(i.e. the egress router in this AS or the border router in the next AS, that will relay the packet
toward the destination) is based on the information exchanged with
neighbors and on a local configuration implementing its routing policy. 

There are two types of BGP sessions that are used to exchange routes
between routers. eBGP sessions are used between routers in different
ASes, while iBGP sessions are used between routers in the same AS. When
a router receives a route on a iBGP or eBGP session, this route has to pass the
input filter to be eligible
in the BGP decision process that selects the best route(s) toward each
destination prefix. The best route(s) selected by this process is(are) then
forwarded on other BGP sessions after passing through an output filter.

The BGP route selection process, implementing routing policies, is made of several criteria (\cite{bgp2,bgp3}):
\begin{itemize}
\item[1)] Prefer routes with the highest local preference which reflects the
  routing policies of the domain;
\item[2)] Prefer routes with the shortest AS-level Path;
\item[3)] Prefer routes with the lowest origin number, e.g., the routes
  originating from IGP are most reliable;
\item[4)] Prefer routes with the lowest MED (multiple-exit discriminator) type
  which is an attribute used to compare routes with the same next AS-hop;
\item[5)] Prefer eBGP-learned routes over iBGP-learned ones 
(referred to as the eBGP$>$iBGP criterion in the sequel);
\item[6)] Prefer the route with the lowest IGP distance to the egress point (i.e. the so-called 
hot-potato, or early exit, criterion);
\item[7)] If supported, apply load sharing between paths. Otherwise, apply a
  domain-dependent tie-breaking rule, e.g., select the one with the lowest
  egress ID.
\end{itemize}

In this paper we will be particularly interested in routes that
are selected using the 6th criterion, which refers to the link weights of the
domain to select the best route toward a destination.

Consider the network of figure \ref{fig:example2nonodes}. Suppose that routes
to $P_1$ are announced by $N_1$ to $R_1$ and $N_2$ to $R_2$ on eBGP
sessions. Suppose that the routes announced by these two routers have
the same attributes (i.e. local-preference, AS-path, origin number and
MED) after passing the input filters of routers $R_1$ and $R_2$ (this
is very frequent in practice for routes that are received from the
same neighboring AS). Suppose also that these two routes are forwarded
by $R_1$ and $R_2$ to $R_0$ on iBGP sessions. Usually the attributes
are not changed when forwarding routes on iBGP sessions. So $R_0$ has
two routes to reach $P_1$ and these two routes are equivalent
w.r.t. criteria 1 to 4. Both are
received on iBGP sessions so are also equivalent w.r.t. the 5th criterion. In
this case $R_0$ will use its IGP distance to $R_1$ and $R_2$ to select
the best route toward $P_1$. We say that this route is
chosen using the hot-potato criterion by router $R_0$. Note that $R_1$
and $R_2$ will directly forward traffic toward this prefix on their interdomain link
using the eBGP$>$iBGP criterion. So we see that prefixes that are
routed via the hot-potato criterion by some routers will be routed according to
the eBGP$>$iBGP criterion by some others and vice-versa.

\begin{figure}[htbp]
  \psfrag{P1}{{\scriptsize $P_1$}}
  \psfrag{P2}{{\scriptsize $P_2$}}
  \psfrag{P3}{{\scriptsize $P_3$}}
  \psfrag{P4}{{\scriptsize $P_4$}}
  \psfrag{N1}{{\scriptsize $$}}
  \psfrag{N2}{{\scriptsize $$}}
  \psfrag{N3}{{\scriptsize $$}}
  \psfrag{N4}{{\scriptsize $$}}
  \psfrag{R1}{{\scriptsize $R_1$}}
  \psfrag{R2}{{\scriptsize $R_2$}}
  \psfrag{R3}{{\scriptsize }}
  \psfrag{R0}{{\scriptsize $R_0$}}
  \psfrag{AS 1}{{\scriptsize $AS1$}}
  \psfrag{AS 2}{{\scriptsize $AS2$}}
  \psfrag{1/1}{{\scriptsize $1/1$}}
  \psfrag{1/2}{{\scriptsize $1/2$}}
  \psfrag{1/3}{{\scriptsize $1/3$}}
  \psfrag{1/4}{{\scriptsize $1/4$ }}
  \psfrag{Engineered AS}{{\scriptsize Engineered AS}}
  \centering
  \includegraphics[width=4.5cm]{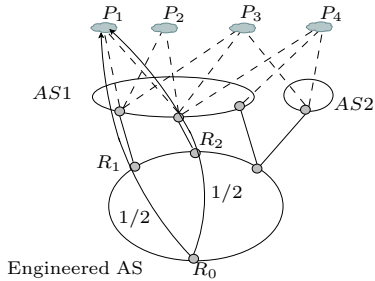}
  \caption{iBGP multipath}
  \label{fig:example2iBGPmultipath}
\end{figure}

Now if $R_1$ and $R_2$ are at the same distance from $R_0$, the 7th 
criterion will be used. By default only one next hop can be chosen
and a tie-break selects the best route. But it is also possible to enable iBGP
multipath load sharing (\cite{bgp2,bgp3}) to balance the load on both paths. As for intradomain
ECMP, a hash table is used to select the particular route of a
packet. Figure \ref{fig:example2iBGPmultipath} supposes that iBGP
multipath is activated and that $R_1$ and $R_2$ are at the same
distance from $R_0$. In this case the traffic going from $R_0$ to
$P_1$ will be split evenly on both paths.

\begin{figure}[htbp]
  \psfrag{P1}{{\scriptsize $P_1$}}
  \psfrag{P2}{{\scriptsize $P_2$}}
  \psfrag{P3}{{\scriptsize $P_3$}}
  \psfrag{P4}{{\scriptsize $P_4$}}
  \psfrag{N1}{{\scriptsize $$}}
  \psfrag{N2}{{\scriptsize $$}}
  \psfrag{N3}{{\scriptsize $$}}
  \psfrag{N4}{{\scriptsize $$}}
  \psfrag{R1}{{\scriptsize $R_1$}}
  \psfrag{R2}{{\scriptsize $R_2$}}
  \psfrag{R3}{{\scriptsize }}
  \psfrag{R0}{{\scriptsize $R_0$}}
  \psfrag{AS 1}{{\scriptsize $AS1$}}
  \psfrag{AS 2}{{\scriptsize $AS2$}}
  \psfrag{1/1}{{\scriptsize $1/1$}}
  \psfrag{1/2}{{\scriptsize $1/2$}}
  \psfrag{1/3}{{\scriptsize $1/3$}}
  \psfrag{2/3}{{\scriptsize $2/3$}}
  \psfrag{1/4}{{\scriptsize $1/4$ }}
  \psfrag{Engineered AS}{{\scriptsize Engineered AS}}
  \centering
  \includegraphics[width=4.5cm]{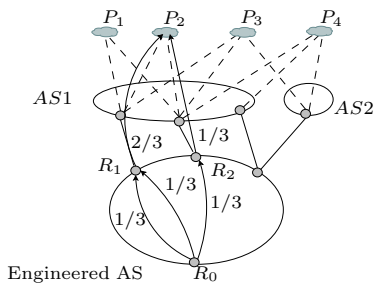}
  \caption{ECMP + iBGP multipath}
  \label{fig:example2ecmpiBGPmultipath}
\end{figure}

If both ECMP and iBGP multipath are activated,
we have to clarify how the traffic is split between multiple
paths. Consider figure \ref{fig:example2ecmpiBGPmultipath}.
Suppose that $R_1$ and $R_2$ are at equal distance from $R_0$. Two
equal cost paths are available from $R_0$ to $R_1$ and only one from
$R_0$ to $R_2$. 
The load sharing implementations in routers we are aware of will send $1/3$ 
of the traffic on each of the 3 available paths at router $R_0$. 

\section{A BGP-aware link weights optimizer}
\label{sec:method}

In this section we present our model of the general traffic
engineering problem. We will use the network of figure
\ref{fig:example2nonodes} to illustrate all the presented
concepts. 

\subsection{Formulation of the traffic engineering problem}

A network is modeled as a directed graph, $G = (N, L)$ whose vertices and edges
represent nodes and links. The basic intradomain topology is composed of all the
nodes and links that belong to the AS. 
We consider two disjoint categories of destination prefixes. 
The \textit{single-egress prefixes} are those prefixes for which the BGP next-hop 
is chosen by one of the first 4 BGP criteria. 
The \textit{hot-potato prefixes} are all the other prefixes. 
For each of them there is at least one router in the domain that has used the hot-potato criterion, 
or a following one, to select the next-hop. 
For each of these \textit{hot-potato prefixes} however, there are also
at least two other routers that forward traffic 
according to the 5th BGP criterion (eBGP$>$iBGP), that has precedence over the
hot-potato criterion (as shown in the example of section
\ref{sec:survey}). The traffic forwarded to the
\textit{single-egress prefixes} constitutes a (hot-potato invariant)
intradomain TM, called  $TM_{invar}$. We also include in that $TM_{invar}$ the
traffic forwarded to the \textit{hot-potato prefixes} originated from
the particular nodes that uses the 5th BGP criterion (eBGP$>$iBGP) to
choose their best route. The remaining traffic forwarded to
\textit{hot-potato prefixes} constitutes $TM_{hp}$.

For every hot-potato prefix we conceptually add a \textit{virtual node} representing it.
Then for every peering link on which equivalent BGP routes (up to criterion 4)
have been announced for that prefix, we extend the intradomain topology 
with a link+node pair representing 
this peering link and the neighboring router on the other side of this link.
Finally we attach all these neighboring routers to the virtual node (representing the hot-potato prefix) 
by adding \textit{virtual links}.
 
Therefore we have three disjoint sets of edges in the topology: $L_{intra}$ is
the set of intradomain links, $L_{inter}$ is the set of
interdomain links, and $L_{virtual}$ is the set of
virtual links. Similarly we split the nodes in the topology into three disjoint sets:
$N_{intra}$ is the set of routers
from the local AS, $N_{neigh}$ is the set of border routers
in neighboring ASes, and
$N_{virtual}$ is the set of virtual nodes.

\begin{figure}[htbp]
  \psfrag{P1}{{\scriptsize $P_1$}}
  \psfrag{P2}{{\scriptsize $P_2$}}
  \psfrag{P3}{{\scriptsize $P_3$}}
  \psfrag{P4}{{\scriptsize $P_4$}}
  \psfrag{N1}{{\scriptsize $N_1$}}
  \psfrag{N2}{{\scriptsize $N_2$}}
  \psfrag{N3}{{\scriptsize $N_3$}}
  \psfrag{N4}{{\scriptsize $N_4$}}
  \psfrag{R1}{{\scriptsize $R_1$}}
  \psfrag{R2}{{\scriptsize $R_2$}}
  \psfrag{R3}{{\scriptsize $R_3$}}
  \psfrag{R0}{{\scriptsize $R_0$}}
  \psfrag{AS 1}{{\scriptsize $AS1$}}
  \psfrag{AS 2}{{\scriptsize $AS2$}}
  \psfrag{1/1}{{\scriptsize $1/1$}}
  \psfrag{1/2}{{\scriptsize $1/2$}}
  \psfrag{1/3}{{\scriptsize $1/3$}}
  \psfrag{1/4}{{\scriptsize $1/4$ }}
  \psfrag{Engineered AS}{{\scriptsize Engineered AS}}
  \centering
  \includegraphics[width=4.5cm]{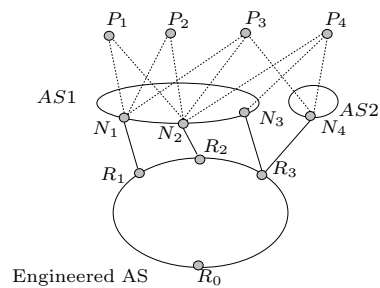}
  \caption{More Complex Topology with virtual nodes}
  \label{fig:example2}
\end{figure}

Figure \ref{fig:example2} shows such a topology. It is the same as
figure \ref{fig:example2nonodes} where prefixes are replaced by
virtual nodes and possible paths to prefixes are replaced by virtual links.
$P_1$, $P_2$, $P_3$ and $P_4$ are HP prefixes that compose $N_{virtual}$. 
The BGP-equivalent routes (up to rule 4) are announced by $N_1$ and $N_2$ for $P_1$ and $P_2$, by $N_1$,
$N_2$ and $N_4$ for $P_3$, and by $N_2$, $N_3$ and $N_4$ for $P_4$.
$L_{inter} = \{R_1-N_1,R_2-N_2,R_3-N_3,R_3-N_4\}$ and $L_{virtual} =
\{N_1-P_1,N_1-P_2,...\}$. $N_{intra} = \{R_*\}$, $N_{neigh} =
\{N_*\}$, and $N_{virtual} = \{P_*\}$.

Each virtual link ($l \in L_{virtual}$) has infinite
capacity $c_l=\infty$ and a fixed weight $w_l=0$. 
Every other link ($l \in L_{intra} \cup  L_{inter}$) has a capacity $c_l$ and a
weight $w_l$. 
Let us note that interdomain and virtual links are directed (toward
the destination prefix) as no transit via a virtual node is
allowed.

The traffic will follow the shortest path(s) 
based on the link weights. If there are multiple equal cost paths,
traffic is considered to be evenly split among them, as shown on figures
\ref{fig:example2ecmp}, \ref{fig:example2iBGPmultipath} and
\ref{fig:example2ecmpiBGPmultipath}.

Once the paths are chosen, we can associate with each link $l$ a load
$l_l$, which is the proportion of traffic that traverses link $l$ 
summed over all pairs of source/destination nodes.
The utilization of a link $l$ is $u_l = l_l/c_l$.

The goal of the LWO is then to find the set of link weights 
that minimizes our network-wide objective function based on the
loads and/or utilizations of intradomain and interdomain links.

\subsection{Aggregating prefixes}

The problem as formulated in the preceding
section is not solvable in practice. Indeed the number of prefixes in the BGP
routing table of an internet router is about 160,000 and so in the worst
case all the prefixes are hot-potato prefixes and about 160,000 nodes would
be added to the intradomain topology (see section \ref{sec:casestudy}
for the actual number of hot-potato prefixes in the operational
network we have studied).
However all prefixes that are reachable through exactly the same set of possible nodes $\in N_{neigh}$ can be
aggregated (e.g., nodes $P_1$ and $P_2$ in
figure \ref{fig:example2} can be merged) as they are indistinguishable 
from an intradomain routing perspective.
This will drastically reduce the number of virtual
nodes. Note that if $n$ is the number of peering links of
the AS, there can still be $2^n$ virtual nodes in the worst case. 
In practice however it is much lower, as explained in
\cite{feamster2}. Indeed routes are often announced with the same parameters on
peering links with the same neighbor AS.
For the operational network we have used as a case study, 
the number of peering links traversed by hot-potato traffic is 18. 
Out of $2^{18}$ possible different combinations of peering
links, only 26 are actually observed!

We can still go one step further by taking the traffic destined for each
aggregated virtual node into account. For example, in our case study we have noticed that no traffic
is sent to 8 of them, and only a very small volume of traffic is sent to 13 others, 
thus leading to 5 nodes receiving 99.94\% of the hot-potato traffic ($TM_{hp}$). 
So we can basically extend the intradomain topology with these 5 virtual nodes 
without really losing accuracy. This is really significant for the practical efficiency of the LWO.
More precisely, using 5 nodes instead of 18 reduced the average computation time of the algorithm from 582 to 140
seconds\footnote{This is the average computation time 
over 14 runs on different TMs with 50 iterations per run. We have used
50 iterations because we have noticed that increasing this number did
not significantly improve the quality of the solution found on this data. These
simulation times are measured on an IBM computer eServer
325 with 2 AMD opteron 2GHz 64 bits processors and 2GB of memory.} 
without decreasing the quality of the provided solutions.
Stated otherwise, the same computational budget would allow us to find 
a better solution (using more iterations) on the smaller topology.

\begin{figure}[htbp]
  \psfrag{P1}{{\scriptsize $P_1$}}
  \psfrag{PT}{{\scriptsize $P_t$}}
  \psfrag{R1}{{\scriptsize $R_1$}}
  \psfrag{RN}{{\scriptsize $R_n$}}
  \psfrag{Src}{{\scriptsize $Src$}}
  \psfrag{Dst}{{\scriptsize $Dst$}}
  \psfrag{FIXED}{{\scriptsize $TM_{invar}$}}
  \psfrag{HP}{{\scriptsize $TM_{hp}$}}
  \centering
  \includegraphics[width=5cm]{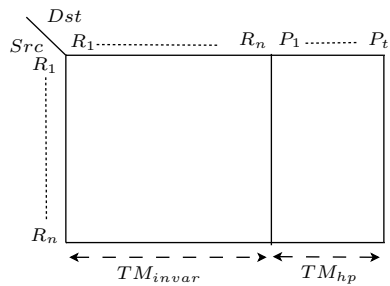}
  \caption{The Aggregated Interdomain Traffic Matrix}
  \label{fig:tm}
\end{figure}

Figure \ref{fig:tm} depicts the structure of the aggregated
interdomain traffic matrix, with one row per edge node in $N_{intra}$
and one column per edge node in $N_{intra}$ or in $N_{virtual}$.

To build this aggregated interdomain traffic matrix we proceed as follows.
Let $(s,p)$ 
be the traffic volume from an ingress node ($s \in N_{intra}$)
to a destination prefix ($p$). If 
$p$ is not a hot-potato prefix (i.e., there is only one
possible egress node $t \in N_{intra}$), we add this traffic volume 
to the pair $(s,t)$ in $TM_{invar}$. If the prefix $p$ is a hot-potato prefix, we
distinguish two subcases. If node $s$ is a possible egress node 
for this prefix, we add this 
traffic volume to the pair $(s,s)$ in $TM_{invar}$ (indeed this traffic
will be routed using the eBGP$>$iBGP criterion). On the other hand, if $s$ is not
one of the possible egress nodes for $p$, 
we add this
amount of traffic to the pair $(s,P_i)$ in $TM_{hp}$, where $P_i
\in N_{virtual}$ is the virtual node associated with the prefix aggregate 
comprising $p$. Now we have our aggregated interdomain traffic matrix,
which is composed of $TM_{invar}$ and $TM_{hp}$.

\subsection{Engineering intra- and interdomain links}
\label{engineeringinter}

LWOs usually try and find nearly optimal set of intradomain link weights.
An optimal set of weights is defined as a set of weights that
associates the minimal value to a predefined objective function. The
objective function is generally the sum over all the links 
of a convex function of the link load and/or utilization. In \cite{fortz1}
they use a piecewise linear convex function of the link utilization and
capacity ($\phi_l$ for link $l$): $\phi = \sum_{l \in L_{intra}} \phi_l$, where $L_{intra}$ is the set of
intradomain links. We reuse this network-wide objective function, while
others could be used in the optimizer.

As interdomain
links are now part of the topology, we can include these links in the
objective function. We are flexible with respect to the inclusion
of these interdomain links in the objective function by adding a
parameter $\alpha$ which determines the relative importance of
interdomain links with respect to intradomain ones. The new
function is $\phi = \sum_{l \in L_{intra}} \phi_l + \alpha
\sum_{l \in L_{inter}} \phi_l$. In section \ref{sec:casestudy} we will compare
cases where $\alpha = 0$ and $\alpha = 1$. Values of $\alpha$ in
  between have not been tested as $\alpha = 1$ seemed to be the good
  compromise in our case. Indeed as shown in section
  \ref{sec:interdomainte} it was possible to engineer interdomain
  links without decreasing the efficiency of the intradomain load
  balance. Note that it could be different in other networks and in
  this case it would be interesting to test other values of $\alpha$.

The inclusion of interdomain links in the objective function is a key
advantage of our method as it allows the LWO to engineer
these interdomain links in addition to intradomain ones. 
With a classical LWO there is no point in including interdomain links 
in the topology and engineer them,
because the intradomain TM used as input pins down the egress node anyway, 
thus assigning the same
load on the interdomain links irrespective of the link weights.
As we relax the constraints on the egress nodes, it seems natural to take advantage of it to
also engineer the traffic on interdomain links. With our method it suffices 
to include these links in the objective function. 

\subsection{Collecting input data for the optimizer}
\label{sec:datacollect}

Our LWO needs as input some information about the
traffic and also some BGP data. The needed traffic
information is the traffic volume from every ingress
router to every destination prefix. For the BGP information we
have to discriminate the hot-potato prefixes from the other ones.
For hot-potato prefixes, we need the set of possible BGP next-hops.
For other prefixes, we just need the unique BGP next-hop.

We will mainly describe the method we
have used in our case study. A monitoring station has been installed inside the network to collect
BGP traces\footnote{We reuse the BGP traces collected for \cite{uhlig-ccr-matrices}.}. It is part of the iBGP full-mesh
and records all the exchanged BGP messages to build BGP traces, 
i.e. daily dumps containing all the routes received by
the monitoring station. In other words, the traces contain for each day all the best routes
used by all the routers of the network toward every possible
destination prefixes.

We distinguish two categories of prefixes:
\begin{itemize}
\item The prefixes for which the same route is selected by all the
  routers as the best route (they will correspond to our earlier definition of single-egress prefixes);
\item The prefixes for which at least two routers in the AS have
selected different best routes (they will correspond to our earlier definition of hot-potato prefixes).
\end{itemize}

The first category of prefixes contains all the prefixes
for which the best route is selected by one of the first 4 criteria
of the BGP process (local preference, AS path, origin
number and MED), 
and the second category contains all the prefixes for
which the best route is selected at a later stage (i.e. by the eBGP$>$iBGP, hot-potato, or tie-break
or load-balancing criteria). Indeed suppose that several routes for the same
prefix are received on different eBGP sessions. If one router selects its best
route by one of the first four criteria, all the other routers will select exactly the
same route by the same criterion, because eBGP data are exchanged "as is" 
on all iBGP sessions and all the routers are part of the iBGP
full mesh. On the other hand if there are at least two equivalent routes 
after the 4th criterion, then each of these routes will be chosen by at least one router 
according to the 5th criterion (eBGP$>$iBGP), namely the border router that has
received that route on its eBGP session.

So we can deduce that if we see only one route for one prefix in the BGP trace,
this means that this prefix is not a hot-potato prefix. 
If this prefix appears at least twice this
means that this prefix is routed by the 5th, 6th or 7th criterion
depending on the router. This prefix is anyway a hot-potato prefix, because
even though some routers have chosen their best route by the 5th criterion,
other routers must have used the 6th or 7th criterion in this case.

\subsection{Incorporating changes in a classical LWO}

We have modified the classical LWO to 
include BGP considerations. Three types of links
(intradomain, interdomain and virtual) are now present in the model. 
Intradomain links are unchanged. Interdomain links have a finite capacity and a weight. 
These are considered in the objective function, weighted by the $\alpha$
parameter.
Finally virtual links have infinite capacities, are not considered in 
the objective function, and have a null weight. 
After these modifications a classical LWO, equipped with all its heuristics, 
can be applied on our extended model.

Notice that the classical LWO considers implicitly that it is 
possible to split the traffic evenly along several equal cost paths.
Therefore it will be necessary to enable ECMP in the network to really
get the expected performance. This is anyway a very reasonable choice.
Moreover, it was shown (in \cite{diot} for the Sprint network)
that ECMP improves robustness. In \cite{diot2} the authors claim that
having multiple shortest paths between pairs of routers provides the
ability to switch over to another path in case of link failure without
overlapping with the previous path of another node, which could have lead
to a transient forwarding loop. It is also said that this is useful
to reduce the latency for forwarding-plane convergence for IGP
routing changes. Similarly to ECMP, we have considered that it is possible to split the traffic
evenly along multiple equal shortest-paths up to the virtual node.
So to get the expected performance the network administrator will have to enable
iBGP multipath load sharing. 
Enabling iBGP multipath load sharing is again a natural choice for traffic engineering and is
easily enabled on routers of main equipment vendors.

\subsection{Respecting the eBGP$>$iBGP criterion}
\label{sec:ebgpibgp}

If next-hop-self is not activated in the network, it is possible to
let the optimizer choose weights on interdomain links. This gives
more knobs to tune to the LWO, in addition to the intradomain links
weights. The pros is that the LWO may potentially find a better
solution, and the cons is the larger search space that increases the
computation time to performance ratio. In large networks it may become
too costly to assign link weights to interdomain links. 

\begin{figure}[htbp]
  \centering
  \includegraphics[width=3cm]{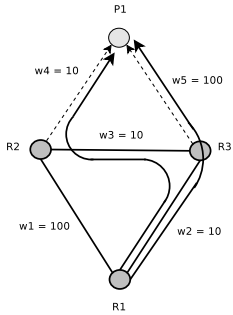}
  \caption{Toy Example - with link weights}
  \label{fig:toyexamplesimplweights}
\end{figure}

Moreover, assigning weights to interdomain links may contradict 
the eBGP$>$iBGP criterion. We explain this point on
the simplified network of figure \ref{fig:toyexamplesimplweights}.
Suppose that the LWO has found the
link weights indicated on the figure. We can easily compute that
the shortest path tree toward destination prefix $P_1$ is $R_1$ - $R_3$ - $R_2$ -
$P_1$. And that is exactly what the LWO has considered during its optimization.
However traffic sent by $R_1$ to $P_1$ will actually follow another path, namely 
$R_1$ - $R_3$ - $P_1$, because according to the eBGP$>$iBGP rule,
which has precedence over the hot-potato rule,
$R_3$ prefers to forward this traffic directly on 
its peering link, although the path via $R_2$ has a lower cost (in terms of weights).

In our simulations we force interdomain link weights to 0, while all
intradomain links are constraint to have integer weights $\ge 1$, so that this
problem is avoided. Indeed for example in the simplified network of
figure \ref{fig:toyexamplesimplweights} the shortest path from $R_3$ to $P_1$ will always be $R_3$ - $P_1$
(weight = 0) and never $R_3$ - $R_2$ - $P_1$ (weight $\ge 1$). Note that
setting all the weights of interdomain links to 0 still allows us to engineer
interdomain links by including these in the objective function as explained
in section \ref{engineeringinter}. So this is not a shortcoming and this is
confirmed by the good results of the simulation study.

\subsection{Simplifying the model}

\begin{figure}[htbp]
  \psfrag{P1}{{\scriptsize $P_1$}}
  \psfrag{P2}{{\scriptsize $P_2$}}
  \psfrag{P3}{{\scriptsize $P_3$}}
  \psfrag{P4}{{\scriptsize $P_4$}}
  \psfrag{N1}{{\scriptsize $N_1$}}
  \psfrag{N2}{{\scriptsize $N_2$}}
  \psfrag{N3}{{\scriptsize $N_3$}}
  \psfrag{N4}{{\scriptsize $N_4$}}
  \psfrag{R1}{{\scriptsize $R_1$}}
  \psfrag{R2}{{\scriptsize $R_2$}}
  \psfrag{R3}{{\scriptsize $R_3$}}
  \psfrag{R0}{{\scriptsize $R_0$}}
  \psfrag{AS 1}{{\scriptsize $AS1$}}
  \psfrag{AS 2}{{\scriptsize $AS2$}}
  \psfrag{1/1}{{\scriptsize $1/1$}}
  \psfrag{1/2}{{\scriptsize $1/2$}}
  \psfrag{1/3}{{\scriptsize $1/3$}}
  \psfrag{1/4}{{\scriptsize $1/4$ }}
  \psfrag{Engineered AS}{{\scriptsize Engineered AS}}
  \centering
  \includegraphics[width=4cm]{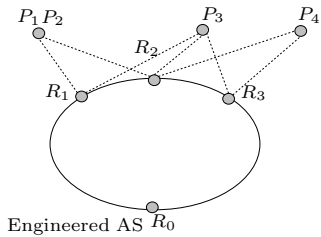}
  \caption{Simplified Model}
  \label{fig:example2-simplified}
\end{figure}

When using the LWO without optimizing 
interdomain links (i.e. only intradomain links are in the objective
function), a simplification of the model is possible. Indeed we can
remove all the interdomain links ($L_{inter}$) and all the neighbor
nodes ($N_{neigh}$) from our model. Figure \ref{fig:example2} would
result in this case in figure \ref{fig:example2-simplified} (where
$P_1$ and $P_2$ have already been aggregated). Indeed in this case the
model has just to include all the possible egress nodes for each
traffic. This simplification decreases the number of links and nodes
of the model and so improves the efficiency of the optimizer.

\section{Simulations on an operational network}
\label{sec:casestudy}

We have tested our algorithm on real data of a multi-gigabit operational network that spreads over the European continent and is composed of about 25 nodes and 40 bidirectional
intradomain links. Link capacities range from 155Mbps to 10Gbps. It is a transit network  
that has two providers connected with about 10 interdomain
links, has other peer ASes connected with about 15 shared-cost links, and has more
than 25 customer ASes, which are mainly single-homed. The total traffic exchanged is 
about 10 Gbps on average.

In this network there is an iBGP full
mesh, MEDs are currently not used, and there are three different 
local preference values: the lowest value is used for routes learned 
from provider links, the intermediate 
value is used for routes learned from shared-cost peering links, and the highest
value is used for routes learned from customer links. Route parameters are
exchanged unmodified on all iBGP sessions. We have used the technique
exposed in section \ref{sec:datacollect} to build our model. We have
used netflow data dumped every 15 minutes on every
ingress router with a sampling rate of $1/1000$, aggregated per
ingress node and destination prefix.
We had access to about one month of traces, one BGP dump per
day and one sampled netflow file for each ingress router. 
With these data we have generated 2,512 aggregated interdomain
traffic matrices (each matrix is an average over 15
minutes). This whole set of traffic matrices is representative of the
traffic on the studied network. Some of these induce a low load on the
network while some induce a high load\footnote{The set of intradomain
  traffic matrices built from the same BGP data and neflow traces is described in \cite{uhlig-ccr-matrices}.}.

The average number of prefixes is 160,973 of which 97.2\% (156,407) are
hot-potato prefixes.
If we now take traffic into account, we
have measured that these 97.2\% amount to 35.6\% of the traffic on average. 
This is still enough to have a significant impact on the link
loads of the network. Over all recorded TMs, the peak value is 51.7\% of the traffic 
and the minimal value is 24.6\%.
Another interesting fact is that on average 
99.94\% of hot-potato traffic is destined for the 5 biggest clusters of prefixes. 
The sets of interdomain links giving access to each of these 5 clusters of
prefixes are either all peering links to a neighboring AS (for 3 clusters), or a mix of peering
links from two such ASes (for 2 clusters).

We have run different versions of the LWO on a large
number of traffic matrices. Section \ref{sec:intradomainte} presents
some simulation results demonstrating the intradomain traffic
engineering capabilities of our algorithm while section
\ref{sec:interdomainte} demonstrates that interdomain traffic
engineering is also possible. All the simulations consider that ECMP and
iBGP multipath are enabled.

\subsection{Intradomain TE}
\label{sec:intradomainte}

\begin{figure}[htbp]
  \centering
  \includegraphics[width=8cm]{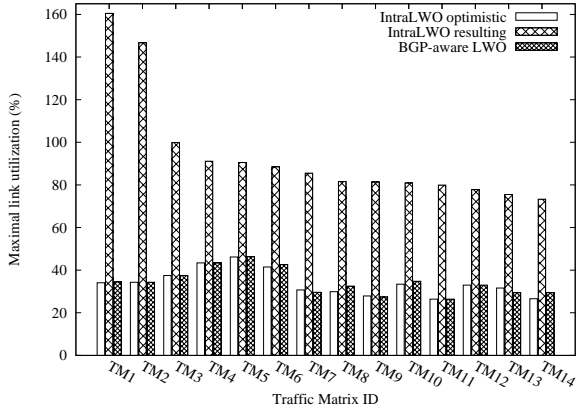}
  \caption{$U_{max}$ values for some worst case TMs}
  \label{tab:resultumax}
\end{figure}

We first compare a classical LWO (denoted \textit{IntraLWO}) with
our BGP-aware optimizer (denoted \textit{BGP-awareLWO}). 
To execute \textit{IntraLWO} we had to generate for each interdomain TM the
corresponding intradomain TM where
the hot-potato traffic is routed considering the present (i.e., non engineered) link
weights. So these intradomain TMs
are those that would be measured in the network.
For the comparison we have run both
optimizers on all the 2,512 aggregated
interdomain TM. Optimizers consider weights in a range from 1 to 150.
Figure \ref{tab:resultumax} shows the maximal intradomain link
utilization ($U_{max}$) for some worst-case TMs. 

We have run \textit{IntraLWO} on every intradomain TM, and computed the 
resulting maximal link utilization, assuming that the intradomain TM remains invariant 
(thus ignoring hot-potato effects). In the sequel these values are denoted
\textit{IntraLWO-optimistic}. For this link weights setting, if
hot-potato effects are taken into account, we get
the resulting maximal intradomain link utilization denoted
\textit{IntraLWO-resulting}. These are the real values that would be
observed if the optimized link weights were installed in the network. 
These values are very different, and sometimes the resulting maximal utilization
is even worse than the routing without link weight optimization (not present in the
figure). Finally we have run our \textit{BGP-awareLWO}
and we can see that the maximal link utilizations are very good. 
Figure \ref{tab:resultumax} shows a selection of TMs providing
the worst-case values for \textit{IntraLWO-resulting}\footnote{
    In this case we define worst case values as values of traffic matrices providing the highest
    intradomain maximal link utilizations.}. The average reduction of $U_{max}$
from \textit{IntraLWO-resulting} to \textit{BGP-awareLWO} over all TMs
is 4.5\%, but let us outline that the worst-case TMs do matter much more, because
the main goal of our LWO is to filter out the unexpectedly bad link weights settings proposed
by a classical LWO.
In all cases the real minimal value of $U_{max}$
achievable \textit{in practice} are the values of \textit{BGP-awareLWO}, since
the \textit{IntraLWO-optimistic} are disqualified in the comparison.

\begin{figure}[htbp]
  \centering
  \includegraphics[width=8cm]{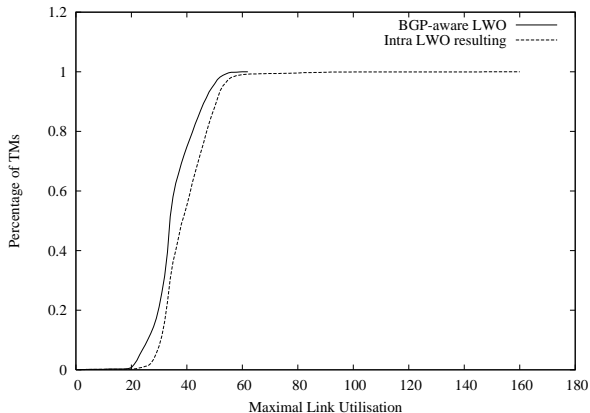}
  \caption{CDFs of $U_{max}$ over all TMs  for \textit{BGP-awareLWO} and \textit{IntraLWO-resulting}}
  \label{fig:umax}
\end{figure}

\begin{figure}[htbp]
  \centering
  \includegraphics[width=8cm]{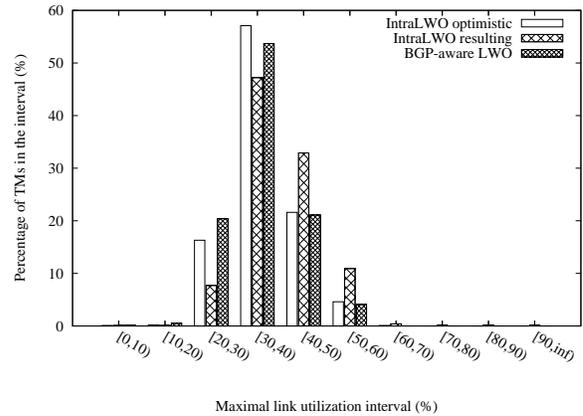}
  \caption{Proportions of TMs in each $U_{max}$ interval}
  \label{tab:resultumax2}
\end{figure}

Figure \ref{fig:umax} shows the CDFs (cumulative distribution functions) of the maximal link utilization over the 2,512 TMs
for \textit{BGP-awareLWO} and
\textit{IntraLWO-resulting}. \textit{IntraLWO-optimistic} is not
depicted on the figure because it would be almost mixed up with \textit{BGP-awareLWO}.
We can clearly see 
that \textit{BGP-awareLWO} is better than \textit{IntraLWO-resulting}. 
Figure \ref{tab:resultumax2} gives the proportions of TMs per range of maximal link utilizations.
In this figure we can see that \textit{BGP-awareLWO} takes advantage of
the freedom of choice of the egress point(s) for hot-potato traffic. Indeed 
\textit{BGP-awareLWO} is slightly better than \textit{IntraLWO-optimistic}.
For example there are 3.4\% less TMs in the [30, 40)
range. This indicates that our optimizer can 
change the egress point of some hot-potato traffic to better engineer
the network. 

Concerning the computational efficiency of the LWO,
adding the virtual links and nodes has roughly doubled the computation
time. We consider that this is not a high cost given the improved quality
of the solutions found.

One may wonder why \textit{BGP-awareLWO} does not always find a
better solution than \textit{IntraLWO-optimistic} (figure  
\ref{tab:resultumax}). It is because the objective
function does not strictly minimize the maximal link utilization 
(i.e., it minimizes the sum over all links of a convex function of the link utilization). 
Therefore even when the solution is slightly better with respect to the objective
function, it can still be a little bit worse with respect to the maximal link utilization.

\subsubsection{In-depth analysis of the worst case scenario}

In this section we would like to analyse the worst case scenario
concerning the maximal link utilization of \textit{IntraLWO-resulting}. 

\begin{figure}[htbp]
  \psfrag{S1}{{\scriptsize $S$}}
  \psfrag{P1}{{\scriptsize $\mathcal{P}_1$}}
  \psfrag{P2}{{\scriptsize $\mathcal{P}_2$}}
  \psfrag{P3}{{\scriptsize $\mathcal{P}_3$}}
  \psfrag{P4}{{\scriptsize $\mathcal{P}_4$}}
  \psfrag{L}{{\scriptsize $L$}}
  \psfrag{D1}{{\scriptsize $D_1$}}
  \psfrag{D2}{{\scriptsize $D_2$}}
  \psfrag{D3}{{\scriptsize $D_3$}}
  \psfrag{VN}{{\scriptsize $VirtualD_4$}}
  \centering
  \includegraphics[width=8cm]{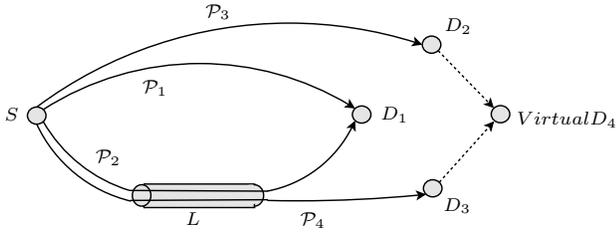}
  \caption{Traffic shifts from one shortest path to another}
  \label{fig:deep-analysis}
\end{figure}

With the worst case traffic matrix, the maximal link utilization is
160\% with the metrics optimized with \textit{IntraLWO}.
The traffic shifts that happen in this case are depicted on figure
\ref{fig:deep-analysis}. If $\mathcal{P}_2 <
\mathcal{P}_1$\footnote{By $\mathcal{P}_2 < \mathcal{P}_1$ we mean
  that the sum of the metrics of the links of $\mathcal{P}_2$ is
  smaller than the sum of the metrics of the links of
  $\mathcal{P}_1$.}, traffic on the flow from $S$ to $D_1$ will be
routed on link $L$, and this will be expected by \textit{IntraLWO}. But if
$\mathcal{P}_4 < \mathcal{P}_3$ while before optimization
$\mathcal{P}_4 > \mathcal{P}_3$, the hot-potato traffic from $S$ to
$VirtualD_4$ will be routed on $L$ and this will NOT be expected by
\textit{IntraLWO}. This situation happens four times on the same low capacity
link\footnote{This link has a capacity of 155 Mbps.} for the
worst case scenario, and for quite big hot-potato traffic
flows compared to the link capacity. 

Before optimization the maximal link utilization is 34.8\%. The
utilization of the problematic link is only 0.4\%. There are only four
intradomain shortest paths that use this link and the total traffic
on these four flows is 0.6 Mbps.

After optimization the problematic link is used in 20
shortest paths instead of 4. This is expected by \textit{IntraLWO} which thinks
that these 20 flows will afford 29 Mbps, leading to a utilization
of only 18.9\% ($<$ 34.1\%, \textit{IntraLWO} thinks that this link is
not the most utilized link). What is not expected by \textit{IntraLWO} is that 4 of these
shortest paths will also attract hot-potato traffic. The hot-potato
traffic which is shifted on one of these shortest paths comes from one
of the 19 remaining shortest paths using this link. So this shift has
no effect on the load of this link. But the hot-potato traffic
attracted on the three remaining shortest paths comes from other flows
whose shortest path does not include the problematic link. These
three flows attract a total amount of 220 Mbps of hot-potato
traffic, which is more than the capacity of the link.

\subsubsection{Increasing the bottleneck links capacities and the traffic matrices}

\begin{figure}[htbp]
  \centering
  \includegraphics[width=8cm]{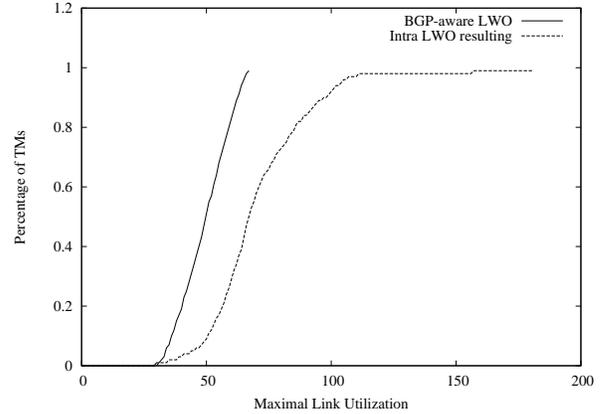}
  \caption{CDFs of $U_{max}$ over all TMs  for \textit{BGP-awareLWO} and \textit{IntraLWO-resulting} for the updated topology}
  \label{fig:cdfUmaxUpdatedTopo}
\end{figure}

To analyse whether the presence of low capacity links has any impact on our results, we did also run our
algorithm on a modified version of the topology, where all the 155Mbps links have
been replaced by 622Mbps links. We have also doubled all the elements of the traffic
matrices in order to reflect a possible increase in the traffic demand in
the future. With this version of the topology and traffic matrices, we have noticed that
the impact of hot-potato reroutings on $U_{max}$ after a LWO optimization is larger than with the
initial topology and load. Indeed the mean reduction of $U_{max}$ over all TMs from \textit{IntraLWO-resulting}
to \textit{BGP-awareLWO} is now 21.8\% instead of 4.5\%. This can be observed
on the CDF of figure \ref{fig:cdfUmaxUpdatedTopo} for the updated topology and
traffic matrices, which should be compared to figure \ref{fig:umax} for
the initial data. We can also observe that for more than 45\% of the traffic matrices,
\textit{IntraLWO-resulting} leads to a $U_{max}$ greater than 67.8\% which is the $U_{max}$
reached on the worst case TM by \textit{BGP-awareLWO}.

Over all TMs $U_{max}$ have been observed on at least 10 different links. There are 7.5\%
of the traffic matrices for which $U_{max}$
is greater than 100\% for \textit{IntraLWO-resulting}, and these high $U_{max}$ values
can be observed on 6 different links, out of which only 2 are 622 Mbps links.
The worst case traffic matrix concerning $U_{max}$ for \textit{IntraLWO-resulting}
induces a utilization of 189.1\% on a link whose capacity is 2.5 Gbps.
These results demonstrate that it is not always the same lowest capacity link that induces
the highest utilization in the network.

We have also analysed CDF curves for the second, third, fourth and fifth most utilized links.
For the second most utilized link, results are similar to those shown on figure 15, with a peak
maximal utilization for \textit{IntraLWO-resulting} reaching 175.9\%, and a maximal utilization
being above 100\% for 2\% of the traffic matrices. Concerning the third most utilized links,
hot-potato reroutings have less disastrous consequences, while still significant in the worst
case as the maximal utilization peaks at 95.3\% for \textit{IntraLWO-resulting} while it
peaks at 62.3\% for \textit{BGP-awareLWO}.

\subsection{Interdomain TE}
\label{sec:interdomainte}

\begin{figure}[htbp]
  \centering
  \includegraphics[width=8cm]{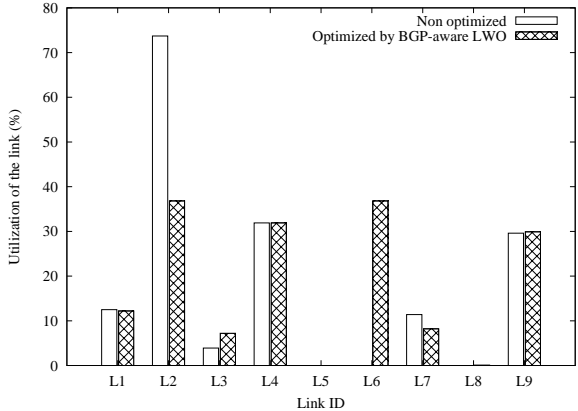}
  \caption{Interdomain link utilizations}
  \label{tab:interdomainte}
\end{figure}

One of the most innovative feature of our LWO is its ability to engineer traffic on the
interdomain links. We first analyse the maximal link utilizations of the
interdomain links with the present link weights. The
average value of Interdomain $U_{max}$ over all TMs is 36.8\%. 
This value can peak at 73.7\%. 
We have selected the worst TMs in this respect\footnote{ Here
    worst case TMs means TMs providing the highest interdomain link
    utilization with present link metrics.} and run {\textit BGP-aware LWO}
on them with interdomain links in the objective function. 
The results are shown in figure \ref{tab:interdomainte} for the peak TM. 
The maximal interdomain link utilization is reduced from 73.7\% to
36.8\% when using {\textit BGP-aware LWO}. It shows that the LWO can take advantage of hot-potato routing
to also engineer traffic on interdomain links.

\begin{figure}[htbp]
  \centering
  \includegraphics[width=8cm]{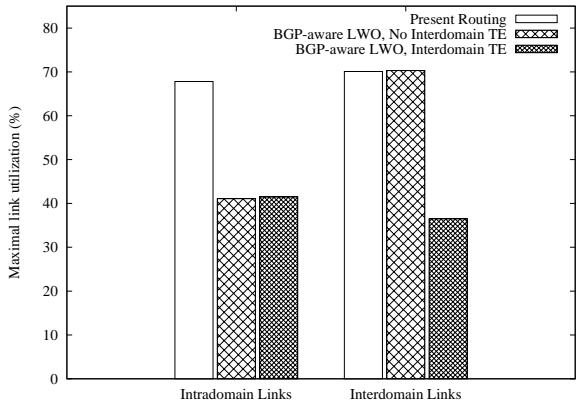}
  \caption{Combined Intra- and Interdomain Traffic Engineering}
  \label{tab:interdomainte2}
\end{figure}

We now show that the optimization of interdomain links
is not done at the expense of intradomain links. 
To this end we have run {\it BGP-aware LWO} with and without
interdomain links in the objective function ($\alpha = 1$ or $\alpha = 0$, see section \ref{engineeringinter}) 
on the 50 TMs
leading currently to the maximal interdomain link
utilization. Figure \ref{tab:interdomainte2} presents the average
intradomain and interdomain $U_{max}$ values for these matrices.
It shows that \textit{BGP-aware LWO} with all links in its
objective function can optimize interdomain links almost without 
impacting intradomain links. The average
intradomain $U_{max}$ value is indeed almost equivalent in both cases.

\section{Future Work}
\label{sec:extensions}

A known potential issue with LWOs is route instability. As there is no mutual agreement
on the egress/ingress points between ASes, it is not guaranteed that
two neighboring ASes (say $AS_x$ and $AS_y$) running their LWO will not oscillate,
one reoptimizing its link weights after the other. Indeed each link weights
optimization in $AS_x$ can lead to a change of some egress points, changing
the traffic matrix in $AS_y$ which may trigger the reoptimization of
the link weights in this AS, and so on, leading to route oscillations.

Such instability may already happen with classical BGP-blind LWOs and, as our BGP-aware
LWO does not address this issue, some instability may also potentially exist.

In \cite{mahajan2005} the authors propose a method to negotiate
the BGP egress point between neighboring ASes. This technique should
remove oscillations provided that it is possible to fix the egress point,
which is not easy in OSPF/ISIS networks. In \cite{mahajan2005} the
authors consider MPLS networks instead.

The related problem of BGP route oscillations when interdomain traffic
engineering techniques are used is considered in
\cite{yang05}, where sufficient conditions are elaborated to guarantee BGP route stability.
Unfortunately, these conditions are not fulfilled in presence of LWOs (be it BGP-aware or not),
because all LWOs take input traffic into account to choose links weights, which in turn
determine egress points for hot-potato prefixes, and thus the corresponding BGP routes.

This problem of potential oscillations is still an open research
topic, and was not the primary goal of this paper.

\section{Conclusion}
\label{sec:conclusion}

We proposed a BGP-aware Link Weight Optimizer (LWO) 
that extends the 
classical (intradomain) LWO to take into account BGP's hot-potato routing principle. 
The optimized link weights, if deployed, will actually give rise to the 
link loads expected by the optimizer, contrary to a classical (intradomain) 
LWO that may lead to unexpectedly high loads on some links when changing 
weights impact the intradomain traffic matrix.
In practice the method only requires to extend the intradomain topology 
with a limited number of virtual nodes and links, which preserves scalability, 
as shown on an operational network used as a case study. The aggregated interdomain
traffic matrix associated with this extended topology 
replaces advantageously the classical intradomain traffic matrix as input to the
LWO.
On this basis, a classical LWO requires only small modifications to be reused on the extended topology, 
and this allows us to reuse all its well-tuned heuristics.

The most innovative key asset of the method is its ability to
optimize traffic on interdomain peering links as well.
We have shown on a case study that it does so very efficiently at almost no extra computational cost, 
while preserving 
the 5th BGP routing criterion stating that eBGP-learned routes should be preferred to iBGP-learned ones.

As for a classical LWO, our method can be extended to more general 
scenarios including several traffic matrices as input and/or possible link failures. 
Note however that an interdomain traffic matrix used as input is likely to be 
already more stable (and thus representative) than intradomain matrices. 
Indeed the interdomain matrix is invariant under all local hot-potato fluctuations, 
e.g. due to failures. This better stability of the interdomain matrix would allow us 
to use a smaller set of representative matrices as input, which in turn would give 
unique link weights settings that are better optimized for each of them.

Even though our method requires additional inputs to build the interdomain traffic matrix
and some more computation power, this pays off, because our BGP-aware LWO clearly
outperforms classical (intradomain) LWO.

\bibliographystyle{abbrv}
\bibliography{sigproc}

\begin{thebibliography}{10}

\bibitem{totem}
http://totem.run.montefiore.ulg.ac.be/.

\bibitem{agarwal05}
S.~Agarwal, A.~Nucci, and S.~Bhattacharyya.
\newblock {Measuring the Shared Fate of IGP Engineering and Interdomain
  Traffic}.
\newblock In {\em Proc. IEEE ICNP}, November 2005.

\bibitem{balon-icon06}
S.~Balon and G.~Leduc.
\newblock {Dividing the Traffic Matrix to Approach Optimal Traffic
  Engineering}.
\newblock In {\em Proceedings of 14th IEEE International Conference on Networks
  (ICON 2006)}, Singapore, 13-15 Sep. 2006.

\bibitem{bgp2}
{BGP Best path selection algorithm}.
\newblock http://www.cisco.com/warp/public/459/25.shtml.

\bibitem{cao}
Z.~Cao, Z.~Wang, and E.~Zegura.
\newblock {Performance of Hashing-Based Schemes for Internet Load Balancing}.
\newblock In {\em {Proceedings of INFOCOM}}, 2000.

\bibitem{selin}
S.~Cerav-Erbas, O.~Delcourt, B.~Fortz, and B.~Quoitin.
\newblock {The Interaction of IGP Weight Optimization with BGP}.
\newblock In {\em Proceedings of ICISP}, Cap Esterel, France, August 2006.

\bibitem{feamster2}
N.~Feamster, J.~Borkenhagen, and J.~Rexford.
\newblock {Guidelines for interdomain traffic engineering}.
\newblock {\em {ACM SIGCOMM Computer Communications Review}}, October 2003.

\bibitem{rexford2}
A.~Feldmann, A.~Greenberg, C.~Lund, N.~Reingold, and J.~Rexford.
\newblock {NetScope: Traffic engineering for IP networks}.
\newblock {\em IEEE Network Magazine}, pages 11--19, March/April 2000.

\bibitem{feldmann}
A.~Feldmann, A.~Greenberg, C.~Lund, N.~Reingold, J.~Rexford, and F.~True.
\newblock {Deriving traffic demands for operational IP networks: Methodology
  and experience}.
\newblock {\em {IEEE/ACM Transactions on Networking}}, pages 265--279, June
  2001.

\bibitem{fortz1}
B.~Fortz and M.~Thorup.
\newblock {Internet Traffic Engineering by Optimizing OSPF Weights}.
\newblock In {\em {Proceedings of INFOCOM}}, pages 519--528, 2000.

\bibitem{fortz2}
B.~Fortz and M.~Thorup.
\newblock {Optimizing OSPF/IS-IS Weights in a Changing World}.
\newblock {\em {IEEE Journal on Selected Areas in Communications}},
  20(4):756--767, 2002.

\bibitem{fortz3}
B.~Fortz and M.~Thorup.
\newblock {Robust optimization of OSPF/IS-IS weights}.
\newblock In {\em {Proceedings of INOC}}, pages 225--230, October 2003.

\bibitem{bgp3}
{Foundry enterprise configuration and management guide}.
\newblock
  http://www.foundrynet.com/services/\\documentation/ecmg/BGP4.html\#17143.

\bibitem{diot}
G.~Iannaccone, C.-N. Chuah, S.~Bhattacharyya, and C.~Diot.
\newblock {Feasibility of IP restoration in a tier 1 backbone}.
\newblock {\em {IEEE Network}}, 18(2), 2004.

\bibitem{mahajan2005}
R.~Mahajan, D.~Wetherall, and T.~Anderson.
\newblock {Negotiation-Based Routing Between Neighboring ISPs}.
\newblock In {\em Proc. NSDI}, 2005.

\bibitem{OSPF}
J.~Moy.
\newblock {OSPF Version 2}.
\newblock {\em RFC 2328}, April 1998.

\bibitem{nucci}
A.~Nucci, B.~Schroeder, S.~Bhattacharyya, N.~Taft, and C.~Diot.
\newblock {IGP Link Weight Assignment for Transient Link Failures}.
\newblock In {\em Proceedings of 18th International Teletraffic Congress
  (ITC)}, September 2003.

\bibitem{rexford1}
J.~Rexford.
\newblock {\em {Handbook of Optimization in Telecommunications}}, chapter
  {Route optimization in IP networks}.
\newblock Springer Science + Business Media, February 2006.

\bibitem{roughan2003}
M.~Roughan, M.~Thorup, and Y.~Zhang.
\newblock {Traffic Engineering with Estimated Traffic Matrices}.
\newblock In {\em Proc. IMC}, 2003.

\bibitem{diot2}
A.~Sridharan, S.~B. Moon, and C.~Diot.
\newblock On the correlation between route dynamics and routing loops.
\newblock In {\em IMC '03: Proceedings of the 3rd ACM SIGCOMM conference on
  Internet measurement}, pages 285--294, New York, NY, USA, 2003. ACM Press.

\bibitem{bgp}
J.~Stewart.
\newblock {\em {BGP4 : Interdomain routing in the Internet}}.
\newblock {Addison Wesley}, 1999.

\bibitem{teixeira3}
R.~Teixeira, N.~Duffield, J.~Rexford, and M.~Roughan.
\newblock {Traffic matrix reloaded: Impact of routing changes}.
\newblock In {\em Proceedings of Passive and Active Measurement}, March/April
  2005.

\bibitem{teixeira1}
R.~Teixeira, A.~Shaikh, T.~Griffin, and J.~Rexford.
\newblock {Dynamics of hot-potato routing in IP networks}.
\newblock In {\em Proceedings of ACM SIGMETRICS}, June 2004.

\bibitem{uhlig-ccr-matrices}
S.~Uhlig, B.~Quoitin, S.~Balon, and J.~Lepropre.
\newblock Providing public intradomain traffic matrices to the research
  community.
\newblock {\em ACM SIGCOMM Computer Communication Review}, 36(1):83--86,
  January 2006.

\bibitem{cope}
H.~Wang, H.~Xie, L.~Qiu, Y.~R. Yang, Y.~Zhang, and A.~Greenberg.
\newblock {COPE: Traffic Engineering in Dynamic Networks}.
\newblock In {\em {Proceedings of ACM SIGCOMM}}, September 2006.

\bibitem{yang05}
Y.~Yang, H.~Xie, H.~Wang, A.~Silberschatz, A.~Krishnamurthy, L.~Yanbin, and
  L.~E. Li.
\newblock On route selection for interdomain traffic engineering.
\newblock {\em IEEE Network}, 19(6):20--27, Nov.-Dec. 2005.

\bibitem{zhang}
Y.~Zhang and Z.~Ge.
\newblock {Finding Critical Traffic Matrices}.
\newblock In {\em Proceedings of the International Conference on Dependable
  Systems and Networks (DSN)}, June 2005.

\end{thebibliography}
\end{document}